\documentclass[useAMS,usenatbib]{mn2e}
\usepackage{graphicx}
\title{Decaying Sterile Neutrinos as a Heating Source in the Milky 
Way Center} 
\author[Chan and Chu]{M.~H.~Chan and 
M.~-C.~Chu \thanks{mhchan@phy.cuhk.edu.hk, mcchu@phy.cuhk.edu.hk}\\
Department of Physics and Institute of Theoretical Physics, The Chinese 
University of Hong Kong\\
Shatin, New Territories, Hong Kong, China}

\begin{document}

\date{Accepted XXXX. Received XXXX} 

\pagerange{\pageref{firstpage}--\pageref{lastpage}} \pubyear{XXXX}

\maketitle

\label{firstpage}

\begin{abstract}
Recent Chandra and Newton observations indicate that 
there are two-temperature components ($T \sim$ 8 keV, 0.8 keV) 
of the diffuse x-rays emitted from deep inside the center of Milky Way. We 
show that this can be explained by the existence of sterile 
neutrinos, which decay to emit photons that can be 
bound-free absorbed by the isothermal hot gas particles 
in the center of Milky Way. This 
model can account for the two-temperature components naturally as 
well as the energy needed to maintain the $\sim$ 8 keV temperature in the 
hot gas. The predicted sterile neutrino mass is between 16-18 keV.
\end{abstract} 

\section{Introduction}
Recently, a large amount of diffuse x-ray data have been obtained by 
Chandra, BeppoSAX, Suzaku and XMM-Newton, giving a complex picture near 
the Milky Way center 
\citep{Muno,Rockefeller,Senda,Hamaguchi,Sidoli,Sakano}. The data 
indicates that there exists a high temperature ($\sim$ keV) hot gas near 
the Milky Way center. \citet{Kaneda} suggested that the existence of 
two-temperature components of the hot gas can explain the observed x-ray 
spectrum. \citet{Sakano} analysed the XMM-Newton data 
to get 1 keV and 4 keV hot gas components in Sgr A East, a supernova 
remnant located close to the Milky Way center. Later 
\citet{Muno} used the data from Chandra to model the temperature of the two components within 
20 pc as 0.8 keV and 8 keV. The temperature of the soft component 
($T_1 \sim$ 0.8 keV) can be explained by 1 percent of kinetic energy by 
one
supernova explosion in every 3000 years near the Milky Way center, 
corresponding to power $\sim 3 \times 
10^{36}$ erg s$^{-1}$ \citep{Muno}. However, the high temperature of the 
hard component ($T_2 \sim$ 8 keV) cannot be explained satisfactorily. The 
power needed to sustain the high temperature is about $10^{40}$ erg 
s$^{-1}$. Moreover, the emission of the hard component is distributed much 
more uniformly than the soft component and the intensity of the two 
components are correlated, which suggest that they are produced by related 
physical processes \citep{Muno}.
In this article, we present a model to explain the two-temperature 
composition, as well as the high temperature and uniform emission of the 
hard component obtained by Chandra within 20 pc. We assume that there are 
sterile neutrinos with rest mass $m_s \ge 16$ keV in the deep 
galactic 
center and their decay photons continuously supply the energy of the hard 
x-ray emission. The sterile neutrino halo is located at the center of a 
20-pc radius gas cloud which is heated by the decayed photons. Most 
photons are absorbed near the center ($\sim 0-1$ pc) of Milky Way 
and the energy is 
subsequently transferred to the surrounding gas. The major 
heating mechanism is the bound-free collisions 
between the decay photons and the ions in the hot gas. In this 
optically thick region, the hot gas particles are in photoionization 
equilibrium. The high metallicity of 
the gas, $\tilde{Z}_{\rm{metal}} \geq 2-3$ solar metallicity, enhances the 
heating rate of 
the gas and provides enough energy to sustain the high temperature of the 
hard component. The energy absorbed in the central region is 
transferred to the outside optically thin region 
($1-20$ pc) by collisions among 
the electrons to share their energy (collisional equilibrium). 
Also, we assume that the soft and hard components are in 
equilibrium with different uniform temperatures and they are bounded 
hydrostatically.

Although the recent MiniBooNE data challenges the LSND result that 
suggests the existence of eV scale sterile neutrinos \citep{Aguilar}, more 
massive sterile neutrinos (eg. keV) may still exist. 
The fact that active neutrinos have rest mass implies that right-handed 
neutrinos should exist which may indeed be massive sterile neutrinos. The 
existence of the sterile neutrinos has been invoked 
to explain many phenomena such as reionization \citep{Hansen}, missing 
mass 
\citep{Dodelson} and the high temperature of the hot gas in clusters 
\citep{Chan}. Therefore, it is worthwhile to discuss the consequences of 
the existence of massive sterile neutrinos, which may decay into light 
neutrinos and photons.

The existence of the small size keV sterile
neutrino halo is first suggested by \citet{Viollier}. The size
of a self-gravitating degenerate sterile neutrino halo depends on $m_s$
and total mass $M_s$:
$R_s=0.0006(M_s/10^6M_{\odot})^{-1/3}(m_s/16~{\rm keV})^{-8/3}$ pc.
Including the contribution of the baryons, the size will be even smaller. 
The size of the sterile neutrino halo is upper bounded by $R_s \le 
0.0005$ pc \citep{Schodel}. This size is very 
small compared to that of a galaxy and therefore the sterile neutrino halo 
will hide deeply inside the 
galactic center \citep{Munyaneza}. Sterile neutrinos may decay into active 
neutrinos and 
become a strong energy source to galaxies and clusters. The 
decays of keV order sterile neutrinos may also help to solve the cooling 
flow problem in clusters \citep{Chan}.

\section{Bound-free absorption model}
The sterile neutrinos at the center will decay into active neutrinos with 
decay rate $\Gamma$ by the following process:
\begin{equation}
\nu_s \rightarrow \nu_a+ \gamma.
\end{equation}
We assume that the energy of decay photons $E_s \approx m_s/2$ is greater 
than 8 keV because 
the energy of each photon must be greater than the energy of each 
electron in the hot gas in order for the latter to gain energy from 
the photons \citep{Chan}. 
The decayed photons come from the volume emission of the entire sterile 
neutrino halo. The distribution of photon energy has a 
characteristic width determined by the Fermi momentum of the sterile 
neutrinos $p_F$, which is very small compared with the rest mass of the 
sterile neutrinos, $p_F/m_sc \sim 10^{-3}$. Therefore, we can approximate 
the photon spectrum as monochromatic with energy $E_s$. 

Since we have not detected any strong lines of such high energy photons 
from the Milky Way center, the optical depth for decayed photons must be 
much greater than 1. Therefore, the total energy 
emitted by the decayed 
photons must be equal to the total energy gained by the electrons 
($\sim 10^{40}$ erg s$^{-1}$):
\begin{equation}
\sum_iN_s \Gamma (E_s-E_i)P_i \ge 10^{40}~{\rm erg~s^{-1}},
\end{equation}
where $N_s$ is the total number of sterile neutrinos, $E_i$ and $P_i$ 
are the ionization potential and probability of photon absorption by 
$i^{th}$ type ions in the hot gas: 
\begin{equation}
P_i= \frac{a_i \sigma_{bf,i}}{\sum_i a_i \sigma_{bf,i}},
\end{equation}
where $a_i$ is the ratio of the number 
of $i^{th}$ type ions to the total number of ions at 
0.8 keV 
temperature (the number density of the soft component is about ten times 
more than that of the hard component if they are in equilibrium). The 
absorption cross section of the $i^{th}$ type ions is
largest for H-like and He-like ions, which is given by \citep{Daltabuit}:
\begin{equation}
\sigma_{bf,i}=10^{-18} \sigma_{th,i} \left[ \alpha_i \left( 
\frac{E_{th,i}}{E_s} \right)^{s_i}+(1- \alpha_i) \left( 
\frac{E_{th,i}}{E_s} \right)^{s_i+1} \right]~{\rm cm^2},
\end{equation}
where $\sigma_{th,i}$, $\alpha_i$, $E_{th,i}$ and $s_i$ are fitted 
parameters of a particular $i^{th}$ type ion. The total effective cross 
section is $\sum_i a_i \sigma_{bf,i}$. The metallicity $\tilde{Z}_{\rm 
metal}$ near the Milky Way 
center can reach $2-3 \tilde{Z}_{\rm metal, \odot}$ (solar metallicity) 
\citep{Sakano}. The metal ions with the largest 
cross section include Si ($\tilde{Z}_{\rm Si}=8.9 \tilde{Z}_{\rm Si, 
\odot}$), S 
($\tilde{Z}_{\rm 
S}=2.7 \tilde{Z}_{\rm S, \odot}$), Ar ($\tilde{Z}_{\rm Ar}=1.8 
\tilde{Z}_{\rm Ar, \odot}$), Ca ($\tilde{Z}_{\rm Ca}=2.5 \tilde{Z}_{\rm 
Ca, \odot}$) and Fe ($\tilde{Z}_{\rm Fe}=3.8 \tilde{Z}_{\rm Fe, 
\odot}$) (see table 1) \citep{Sakano}. Assuming $\tilde{Z}_{\rm metal}=3 
\tilde{Z}_{\rm metal, \odot}$ 
for other metals and $\Gamma=(3-6) \times 
10^{-19}$ s$^{-1}$ (the best fit value that can solve the cooling flow 
problem) 
\citep{Chan}, we can obtain $N_s$ and the total mass of the sterile 
neutrino halo ($M_s=N_sm_s$) by Eq.~(2) for different $E_s$ (see 
table 2). The value of $M_s$ obtained ($\sim 10^5 M_{\odot}$) is far below 
the upper bound of the 
total mass near the Milky Way center ($2.6 \times 10^{6}M_{\odot}$) 
\citep{Schodel}.

\begin{table*}
 \caption{Metal abundances we have used in the model.}
 \label{table1}
 \begin{tabular}{@{}lcccc}
  \hline
  Element & Atomic number $Z$ & Metallicity $\tilde{Z}_{\rm metal}$ 
  & $a_i$($10^{-4}$)(H-like) & $a_i$($10^{-4}$)(He-like)\\ 
  \hline
  C &6 &3 &6.9 &2.7\\
  N &7 &3 &1.7 &0.84\\
  O &8 &3 &16 &9.6\\
  Ne &10 &3 &2.5 &1.9\\
  Mg &12 &3 &1.0 &0.89\\
  Si &14 &8.9 &3.0 &2.8\\
  S &16 &2.7 &0.43 &0.42\\
  Ar &18 &1.8 &0.066 &0.066\\
  Ca &20 &2.5 &0.056 &0.056\\
  Fe &26 &3.8 &1.8 &1.8\\
  \hline
 \end{tabular}
\end{table*}

To constrain the value of $E_s$, we can consider the upper limit of x-ray 
luminosity in the $2-10$ keV band near the Milky 
Way center, which is $3 \times 
10^{36}$ erg s$^{-1}$ \citep{Muno}. We can obtain the lower limit of the 
optical depth $\tau_{\rm lower}$ by:
\begin{equation}
N_s \Gamma E_se^{-\tau} \le 3 \times 10^{36}~ {\rm erg~s^{-1}},
\end{equation}
where $\tau$ is the optical depth for decayed photons which is given by:
\begin{equation}
\tau= \int n(r) \sum_ia_i \sigma_{bf,i}dr.
\end{equation}
A higher energy of the decayed photons will result in a smaller cross 
section and optical depth. The mass density profile of Milky Way 
center 
can be modeled by a power law $\rho \sim r^{-1.8}$ \citep{Schodel}. For 
simplicity, we assume that the number density of hot 
gas near the center has an isothermal profile:
\begin{equation}
n(r)=n_0 \left( \frac{r}{\rm 1~pc} \right)^{-2}.
\end{equation}
The average number density within 20 pc of the center is about 1 cm$^{-3}$ 
which corresponds to $n_0=133$ cm$^{-3}$. In table 2, we can see that 
$\tau< \tau_{\rm lower}$ 
for $E_s \ge 10$ keV. Therefore, only $E_s=8-9$ keV can be possible for 
decaying sterile neutrinos as the heating source (see Fig.~(1)).  

\begin{figure*}
\vskip5mm
 \includegraphics[width=84mm]{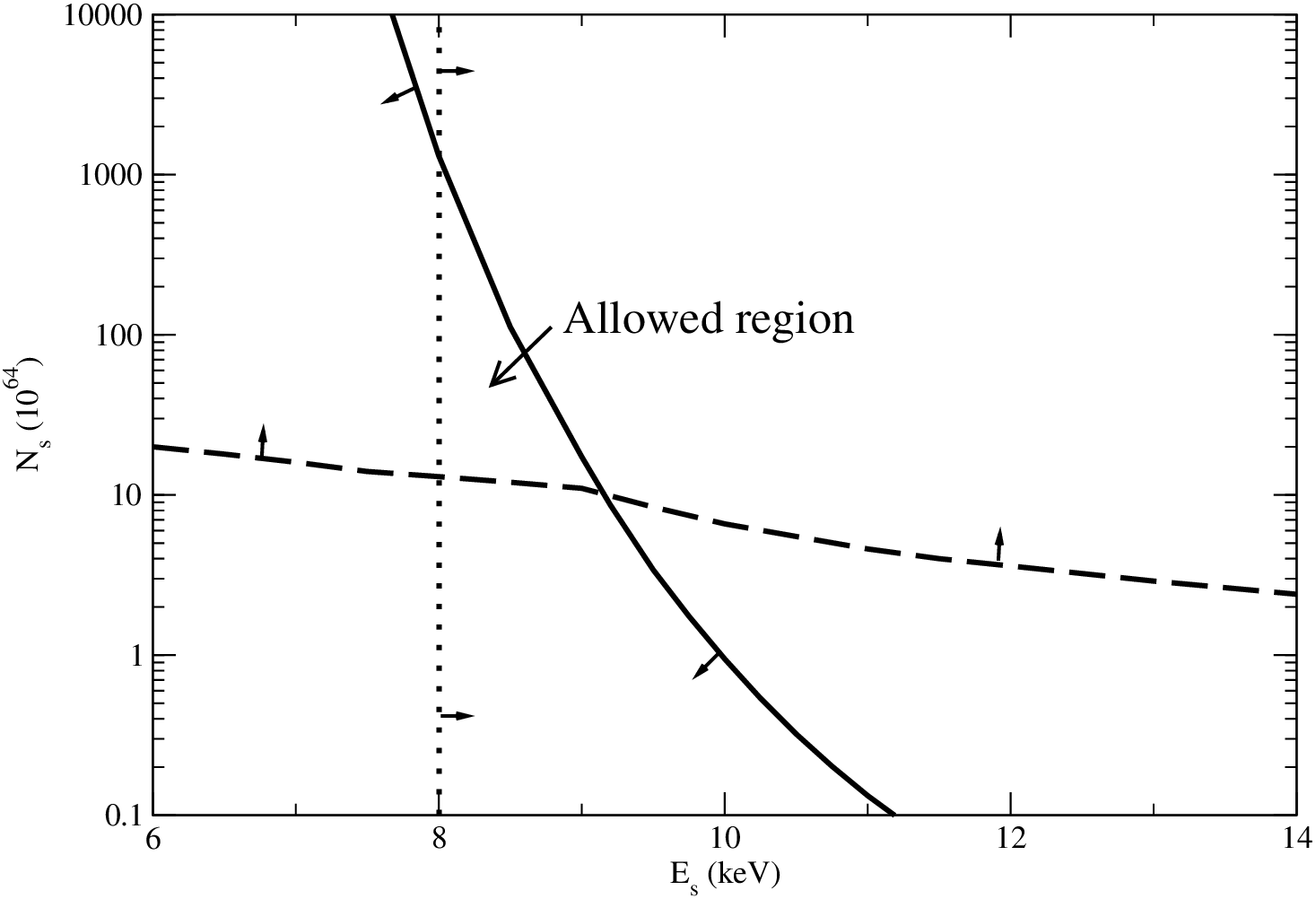}
 \caption{The allowed $N_s$ and $m_s$ by using the constraints in Eq.~(2) 
(dashed line), Eq.~(5) (solid line) and by the temperature of the hard 
component ($E_s \ge 8$ keV)(dotted line).} 
\end{figure*} 

\begin{table*}
 \caption{The values of $N_s$, $M_s$ and $\tau$ for $E_s=8-14$ keV.}
 \label{table2}
 \begin{tabular}{@{}lccccc}
  \hline
  $E_s$ (keV) & $N_s$($10^{66}$) & $M_s$($10^5M_{\odot}$) &$\sum_i 
  a_i \sigma_i$($10^{-23}$~cm$^2$) &$\tau$ &$\tau_{\rm lower}$ \\ 
  \hline
  8 &13 &1.9 &1.77 &14.6 &10.0\\
  9 &11 &1.7 &1.26 &10.4 &9.93\\
  10 &9.1 &1.6 &0.921 &7.61 &9.87\\
  11 &4.6 &0.9 &0.685 &5.74 &9.30\\
  12 &3.6 &0.76 &0.536 &4.43 &9.12\\
  13 &2.9 &0.67 &0.422 &3.48 &8.90\\
  14 &2.4 &0.61 &0.338 &2.79 &8.90\\
  \hline
 \end{tabular}
\end{table*}

Given this power input ($10^{40}$ erg s$^{-1}$), we can estimate the 
equilibrium temperature of the gas at the center of Milky Way. The 
ionizing photons are continuously supplied and the heating rate is quite 
constant as the decay of sterile neutrinos is a slow process (half-life of 
order Hubble time). The energy absorbed 
will then be transferred to the other gas particles mainly
by conduction. The power loss by Bremsstrahlung 
radiation of a hot gas with temperature $T$ is $\Lambda=1.4 \times 
10^{27}n_i^2T^{1/2}$ erg s$^{-1}$ cm$^{-3}$, and the power by its 
adiabatic expansion is 
\begin{equation}
\dot{W}=PV^{2/3}c_s=PV^{2/3} \left(\frac{ \gamma kT}{m_g} \right)^{1/2},
\end{equation}
where $P$, $V$, $c_s$, $\gamma$ and $m_g$ are pressure, volume, sound 
speed, adiabatic index of the hot gas and mean mass of the gas particles 
respectively. The energy loss due to Bremsstrahlung is negligible, being 
only 0.3 percent of the adiabatic cooling \citep{Muno}. We can therefore 
solve for the equilibrium temperature of the gas using Eq.~(8). The 
temperature maintained in this process is given by 
\begin{equation}
T \approx V^{-4/9} \left( \frac{\dot{W}}{n_tk} \right)^{2/3} \left( 
\frac{m_g}{\gamma k} \right)^{1/3}.
\end{equation}
If $\dot{W}=10^{40}$ erg s$^{-1}$ within $20$ pc, then $T \approx 8$ 
keV. In fact, the situation is similar to energy absorption in x-ray 
emitting nebulae. However, in a normal x-ray nebula, the x-ray source is 
usually far away from the gas cloud, whereas in our model the source is 
located at the center of the gas cloud. Furthermore, the energy absorption 
in a normal x-ray nebula is about $10^{30}$ erg s$^{-1}$ so that the 
temperature of the nebula is about eV order \citep{Leahy}. In our model, 
the energy absorption is $10^{40}$ erg s$^{-1}$, which can maintain 
the temperature of the hard component at 8 keV.

\section{Two-temperature components}
When the gas particles are in thermal equilibrium, 
\begin{equation}
\tilde{\Gamma}(n_g,T)= \tilde{\Lambda}(n_g,T), 
\end{equation}
where $n_g$ is the number density of the gas particles, $\tilde{\Gamma}$ 
and $\tilde{\Lambda}$ are the heating and 
cooling rates respectively. It is possible to have multi-temperature 
components in the hot gas, because there may be more than one solutions 
($n_g,T$) satisfying Eq.~(10). The criterion for stability of the solution 
is \citep{Bowers}:
\begin{equation}
\frac{\partial}{\partial T}(\tilde{\Gamma}-\tilde{\Lambda})+ 
\frac{\partial}{\partial \rho}(\tilde{\Gamma}-\tilde{\Lambda})\left( 
\frac{\partial \rho}{\partial T} \right)_P < 0,
\end{equation}
where $\rho$ and $P$ are the mass density and pressure of the hot gas. 
Since the relaxation time of electron-electron collisions is less than 
that of ion-electron collisions, the temperature of electrons $T_e$ may be 
different from the temperature of ions $T_i$ in general. The total heating 
rate inside the gas cloud ($r \le R \approx 20$ pc) is given by: 
\begin{equation}
\tilde{\Gamma} \sim \tilde{\Gamma_0}n_g \sum_i a_i \sigma_{bf,i},
\end{equation}
where $\tilde{\Gamma_0}$ is a constant that depends on the size of the 
region.
The cooling rate of x-ray emitting hot gas by Bremsstrahlung 
radiation is:
\begin{equation}
\tilde{\Lambda} \sim \tilde{\Lambda_0}n_en_g \sum_i 
a_iZ_i^2e^{-2I_i/kT_i}T_e^{1/2},
\end{equation}
where $\tilde{\Lambda_0}$ is a 
constant, $I_i$ and $Z_i$ are the ionization energy and charge of the 
$i^{th}$ type ions respectively. 
Since the entire gas cloud is optically thin for $r \sim$ pc  
and heavy metal 
ions are concentrated at the central region \citep{Sakano}, the bound-free 
absorption of the broad-band 
Bremsstrahlung photons is suppressed ($\tau \le 1$) in a large fraction of 
the volume of the cloud, so that most of 
the Bremsstrahlung photons can escape the gas cloud easily. When the hot 
gas particles are in thermal equilibrium, $\tilde{\Gamma}= 
\tilde{\Lambda}$, and the pressure is given by: 
\begin{equation}
P=P_0 \frac{T_{{\rm keV}}^{1/2} \sum_ia_i \sigma_{bf,i}}{ 
\sum_ia_iZ_i^2e^{-2I_i/kT_i}},
\end{equation}
where $P_0=(1~{\rm keV})^{1/2}k \tilde{\Gamma_0}/ \tilde{\Lambda_0}$ and 
$T_{{\rm keV}}=T_e/(1~ \rm keV)$.

\begin{figure*}
\vskip8mm
 \includegraphics[width=84mm]{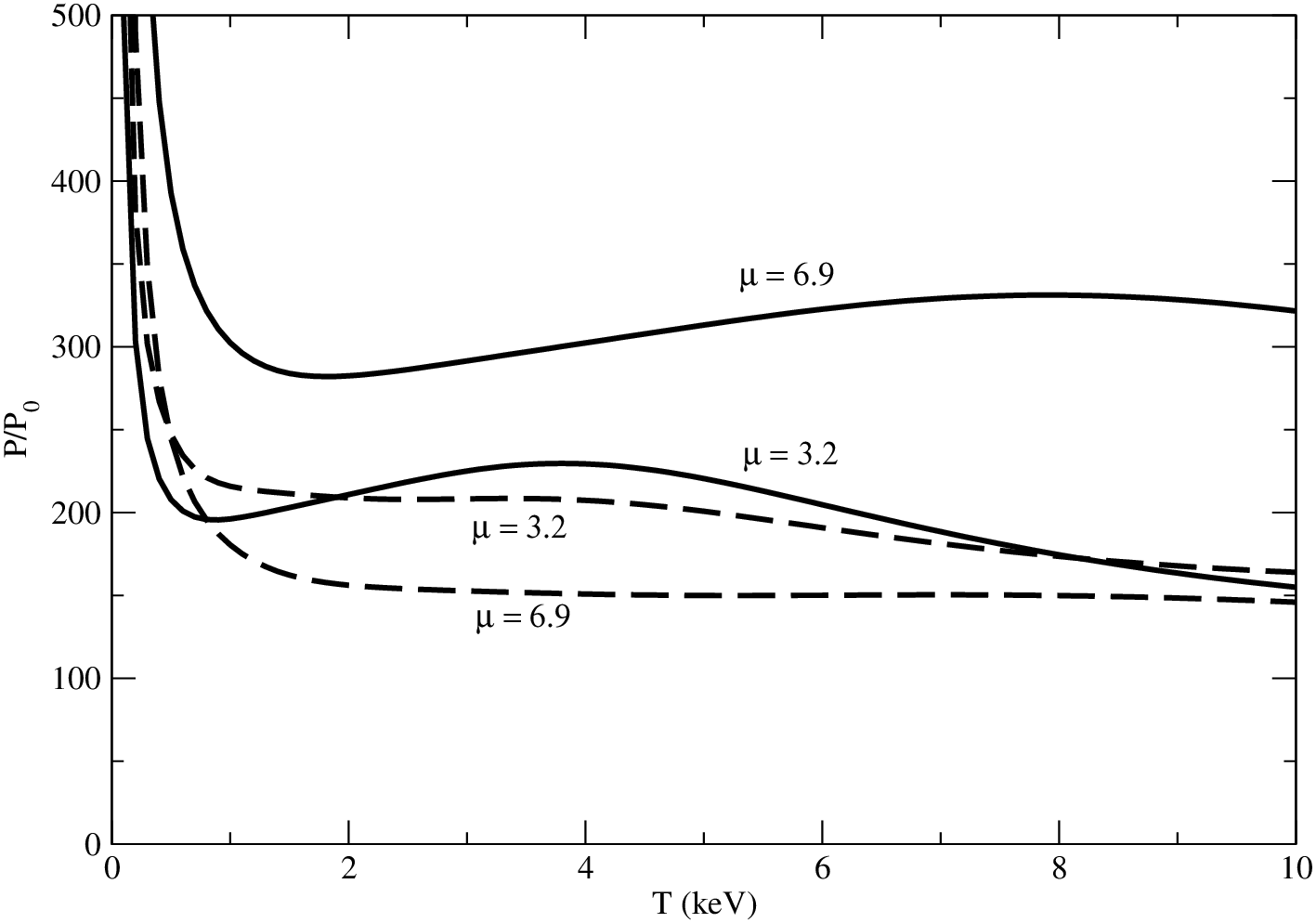}
 \caption{Pressure ($P$) versus temperature ($T$) of the hot gas, given by 
 Eq.~(14) for $\tilde{Z}_{Fe}=0.71$ (solid lines) and 
 $\tilde{Z}_{Fe}=0.14$ (dashed lines) for two different values of $\mu$.}
\end{figure*} 

Fig.~(2) shows the relation between $P$ and $T_e$ by using the mean 
metallicity within 20 pc \citep{Muno}. We notice that for some 
values of $P$ there are two different values of $T_e$.
By putting the heating and cooling rates into Eq.~(11), we get the 
criterion for stable solutions:
\begin{equation}
\frac{\sum_ia_i \sigma_{bf,i}}{\sum_ia_iZ_i^2e^{-\beta_i}} 
\left[ \sum_ia_iZ_i^2e^{-\beta_i} \left( \frac{3}{2}- \beta_i \right) 
\right]< \sum_i \sigma_{bf,i} \left[a_i-T_e \frac{da_i}{dT_i} 
\right],
\end{equation}
where $\beta_i=2I_i/kT_i$. We impose a free parameter 
$\mu$ such that $T_e= \mu T_i$. We define $A$ and $B$ to be the 
expressions on the left and right hand sides of the inequality in Eq.~(15) 
respectively. In Fig.~(3), we
plot $A-B$ against $T_e$ for two values of $\mu$. When 
$\mu =3.2$ and $\mu =6.9$, 0.8 keV $<T_e<$ 4 keV and 1.8 keV 
$<T_e<$ 8 keV are 
unstable solutions respectively as $A-B>0$. We also indicate the unstable 
regions in Fig.~(3), and we notice that a two-temperature phase may 
exist with $T_e \ge 8$ keV and $T_e \le 0.8$ keV for $3.2 \le \mu \le 
6.9$. In Fig.~(2), we can see 
that if there are less large-$Z$ metal ions (especially iron) in 
the hot gas, more 
hot gas particles will shift to the lower temperature phase and only one 
equilibrium solution may be obtained. Clearly, our model 
can account for the two-temperature structure of the hot gas near the 
Milky Way center.

\begin{figure*}
\vskip10mm
 \includegraphics[width=84mm]{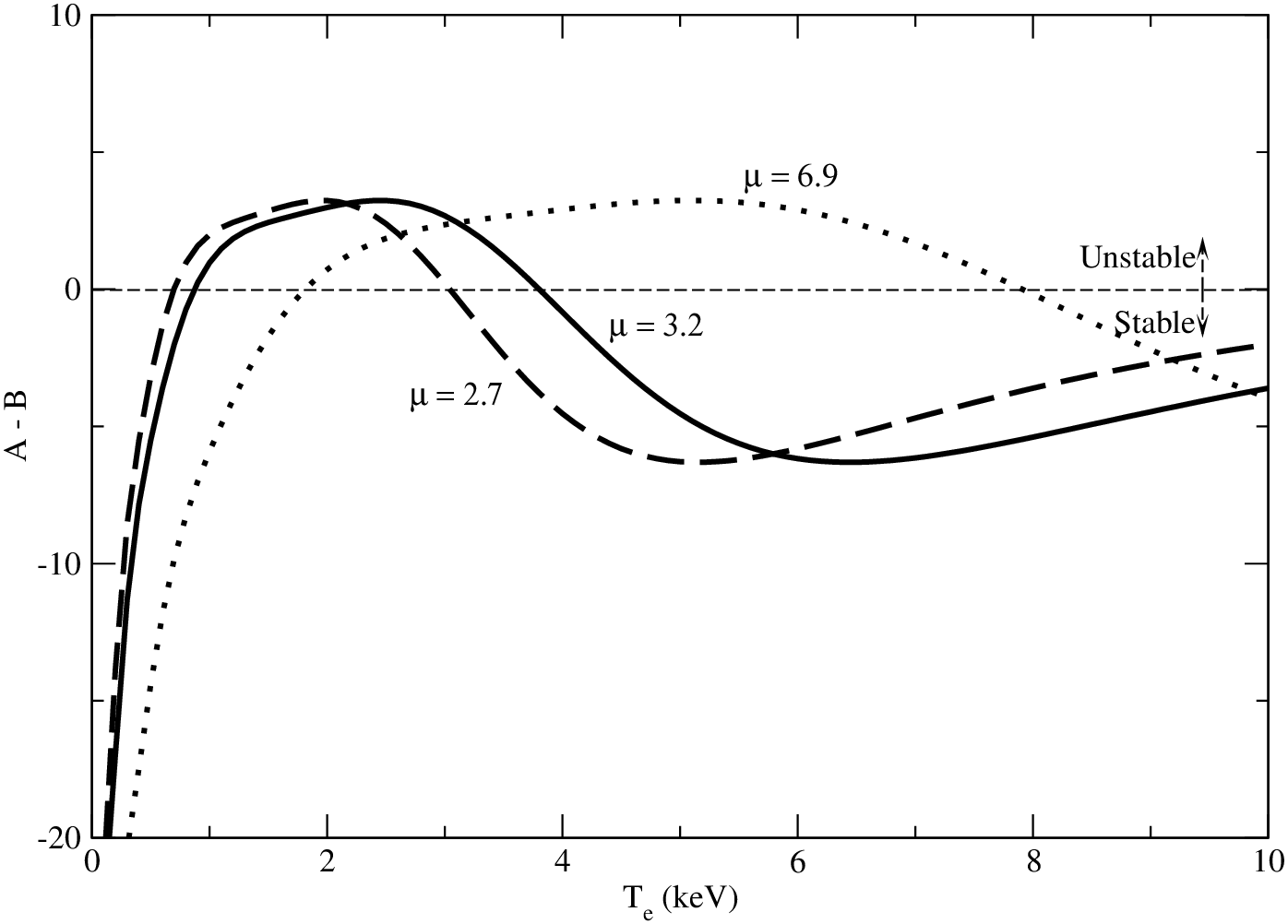}
 \caption{The difference between the left hand side ($A$) and 
 right hand side 
 ($B$) of Eq.~(15) versus $T_e$ for $\mu=2.7$ (dashed line), $\mu=3.2$ 
 (solid line) and $\mu=6.9$ (dotted line).} 
\end{figure*}

\section{discussion}
To explain the origin of the hard component, \citet{Muno}
make use of magnetic 
reconnection driven by the turbulence that supernovae generate in 
the interstellar medium. Magnetic reconnection can heat the hot gas to 
$kT \sim B_{\rm center}^2/8 
\pi n_g$. For $n_g \sim 0.1$ cm$^{-3}$, $B_{\rm center} \sim 0.2$ mG, $kT 
\sim 
8$ keV. However there is not enough evidence to support whether
this mechanism can maintain the high temperature of the hard component 
\citep{Muno}. 

In our model, we have assumed that there exists a sterile neutrino halo 
with $m_s=16-18$ keV in the Milky Way center, which decay to emit $\gamma$ 
with life-time of cosmological order. It provides a large amount 
of energy to the 
hot gas and maintains the extremely high temperature. The bound-free 
collisions provide 
enough energy to the two different temperature components and maintain 
their temperatures. At the same time, a stable two-temperature structure 
in the hot gas can be explained by this heating mechanism naturally. The 
uniform emission of 
the soft and hard components suggests that 
they may come from similar physical processes \citep{Muno}. In our model, 
both components indeed share the same source of energy - the 8-9 keV 
photons emitted by the decays of sterile neutrinos. In our model, the 
sterile 
neutrinos may not be the major component of dark matter. Therefore, any 
bounds on $m_s$ assuming they are the major dark matter candidate does not 
constrain our model severely. The heating rate in 
the Milky Way center is time dependent as there is a decreasing 
number of sterile neutrinos. Therefore, if two galaxies have similar 
chemical compositions, the heating rate is greater for 
the large redshift one, which has more particles in the higher 
temperature component. We therefore predict that the hard component of the 
x-rays 
would be stronger for large redshift and metal-rich galaxies. Moreover, if 
a galaxy has lower metallicity, then only a single temperature component 
may be observed instead of two (see Fig.~(2)). 

\section{acknowledgements}

This work is partially supported by a grant from the Research Grant 
Council of the Hong Kong Special Administrative Region, China (Project No. 
400805).

\label{lastpage}


\begin{thebibliography}{}
\bibitem[Aguilar-Arevalo et al.(2007)]{Aguilar} Aguilar-Arevalo, 
A.~A.~{\it et al.} 2007, Phy.~Rev.~Lett., {\bf 98}, 231801.
\bibitem[Almy et al.(2000)]{Almy} Almy, R.~C.~{\it et al.} 2000, ApJ, {\bf 
545}, 290.
\bibitem[Bowers and Deeming(1984)]{Bowers} Bowers, R.~and Deeming, 
T.~1984, {\it Astrophysics}, Jones and Bartlett, 
London.
\bibitem[Chan and Chu(2007)]{Chan} Chan, M.~H.~and Chu, M.~-C.~ 2007, ApJ, 
{\bf 658}, 859.
\bibitem[Daltabuit and Cox(1972)]{Daltabuit} Daltabuit, E.~and Cox, 
D.~P.~1972, ApJ, {\bf 177}, 855.
\bibitem[Dodelson and Widrow(1994)]{Dodelson} Dodelson, S and Widrow, 
L.~M.~ 1994, Phys.~ Rev.~ Lett., {\bf 72}, 17.
\bibitem[Frogel et al.(1999)]{Frogel} Frogel, J.~A., Tiede, G.~P.~and 
Kuchinski, L.~E.~1999, AJ, {\bf 117}, 2296.
\bibitem[Hamaguchi et al.(2007)]{Hamaguchi} Hamaguchi, K.~{\it et al.} 
2007, astro-ph/0704.0346 v1.
\bibitem[Hansen and Haiman(2004)]{Hansen} Hansen, S.~H.~and Haiman, Z. 
2004, ApJ, {\bf 600}, 26.
\bibitem[Kaneda et al.(1997)]{Kaneda} Kaneda, H.~{\it et al.} 1997, ApJ, 
{\bf 491}, 638.
\bibitem[Leahy et al.(1994)]{Leahy} Leahy, D.~A., Zhang, C.~Y.~and Kwok, 
S.~1994, ApJ, {\bf 422}, 205.
\bibitem[Mapelli and Ferrara(2005)]{Mapelli} Mapelli, M.~ and Ferrara, 
A.~2005, MNRAS, {\bf 364}, 2.
\bibitem[Muno et al.(2004)]{Muno} Muno, M.~P.~{\it et al.} 2004, ApJ, {\bf 
613}, 326.
\bibitem[Munyaneza and Viollier(2002)]{Munyaneza} Munyaneza, F.~and 
Viollier, R.~D.~2002, ApJ, {\bf 564}, 274.
\bibitem[Park et al.(2003)]{Park} Park, S.~{\it et al.} 2003, ApJ, {\bf 
603}, 548.
\bibitem[Rockefeller et al.(2004)]{Rockefeller} Rockefeller, G.~{\it et 
al.} 2004, ApJ, {\bf 604}, 642.
\bibitem[Sakano et al.(2004)]{Sakano} Sakano, M.~{\it et al.} 2004, MNRAS, 
{\bf 350}, 129.
\bibitem[Sch\"{o}del et al.(2002)]{Schodel} 
Sch\"{o}del, R.~{\it{et al.}} 2002, Nature, {\bf 419}, 694.
\bibitem[Senda et al.(2002)]{Senda} Senda, A., Murakami, H.~and Koyama, 
K.~2002, ApJ, {\bf 565}, 1017.
\bibitem[Sidoli et al.(1999)]{Sidoli} Sidoli, L.~{\it et al.} 1999, ApJ, 
{\bf 525}, 215.
\bibitem[Viollier et al.(1993)]{Viollier} Viollier, R.~D.~, Trautmann, 
D.~and Tupper, G.~B.~1993, Phys.~Lett.~B., {\bf 306}, 79.
\bibitem[Zeilik and Gregory(1998)]{Zeilik} Zeilik, M.~and Gregory, 
S.~A.~1998, {\it Introductory Astronomy and 
Astrophysics}, Saunders College Publishing.

\end{thebibliography}
\end{document}